# Seismic Monitoring of the Sun's Far Hemisphere:
# A Crucial Component in Future Space Weather Forecasting

*Principal Author:* Kiran Jain
National Solar Observatory, 3665 Discovery Dr., Boulder, CO 80303, USA

*Co-authors:* C. Lindsey[1], E. Adamson[2], C. N. Arge[3], T. E. Berger[4], D. C. Braun[1], R. Chen[5], Y. M. Collado-Vega[3], M. Dikpati[6], T. Felipe[7], C. J. Henney[8], J. T. Hoeksema[5], R. W. Komm[9], K. D. Leka[1], A. R. Marble[10,2], V. Martinez Pillet[9], M. Miesch[10,2], L. J. Nickisch[11], A. A. Pevtsov[9], V. J. Pizzo[2], W. K. Tobiska[12], S. C. Tripathy[9], J. Zhao[5]

[1]NorthWest Research Associates, Boulder, CO 80301
[2]NOAA/Space Weather Prediction Center, Boulder, CO 80305
[3]NASA/Goddard Space Flight Center, Greenbelt, MD 20771
[4]SWx TREC, University of Colorado, Boulder, CO 80303
[5]Stanford University, Stanford, CA 94305-4085
[6]High Altitude Observatory, NCAR, Boulder, CO 80301
[7]Instituto de Astrofísica de Canarias, La Laguna, Tenerife, Spain
[8]AFRL Space Vehicles Directorate Kirtland AFB, NM 87117
[9]National Solar Observatory, Boulder, CO 80303
[10]CIRES, University of Colorado, Boulder, CO 80303
[11]NorthWest Research Associates, Monterey, CA 93940
[12]Space Environment Technologies, Pacific Palisades, CA 90272-2844

**Synopsis***:* The purpose of this white paper is to put together a coherent vision for the role of helioseismic monitoring of magnetic activity in the Sun's far hemisphere that will contribute to improving space weather forecasting as well as fundamental research in the coming decade. Our goal fits into the broader context of helioseismology in solar research for any number of endeavors when helioseismic monitors may be the sole synoptic view of the Sun's far hemisphere. It is intended to foster a growing understanding of solar activity, as realistically monitored in both hemispheres, and its relationship to all known aspects of the near-Earth and terrestrial environment. Some of the questions and goals that can be fruitfully pursued through seismic monitoring of farside solar activity in the coming decade include:

- What is the relationship between helioseismic signatures and their associated magnetic configurations, and how is this relationship connected to the solar EUV irradiance over the period of a solar rotation?
- How can helioseismic monitoring contribute to data-driven global magnetic-field models for precise space weather forecasting?
- What can helioseismic monitors tell us about prospects of a flare, CME or high-speed stream that impacts the terrestrial environment over the period of a solar rotation?
- How does the inclusion of farside information contribute to forecasts of interplanetary space weather and the environments to be encountered by human crews in interplanetary space?

Thus, it is crucial for the development of farside monitoring of the Sun be continued into the next decade either through ground-based or space-borne observations.

1. **Background**

Our space environment is affected by violent solar episodes that impact modern technological society in numerous ways. For example, electrical power grids, telecommunications, air transport, space activities, navigation, etc., all have been adversely affected by events on the Sun. These occur on time scales from minutes to months, or even years. The origin of these episodes lies in the complex system of magnetic flux that permeates the solar interior, emerges through the photosphere and finally moves into the Sun's outer atmosphere. This magnetic flux has considerable free energy which, released in the aforesaid episodes, drives flares, coronal mass ejections (CMEs), bursts of solar energetic particles (SEPs) and high-speed streams in the solar wind ([Temmer 2021](#)). These anomalies rain into the near-Earth environment, radically impacting it and the interplanetary space environment. As in terrestrial weather, we can do little to change "space weather" as yet. However, there are ways we can reduce its impact through advance warnings, and this depends crucially on how well we understand the physics of solar eruptive phenomena and our ability to forecast space weather.

2. **What We Lose by Observing the Earth-facing Side Only**

While SEPs emanating from energetic eruptions beyond the west limb occasionally have severe impacts on the near-Earth environment, solar eruptive events originating from the far hemisphere near the east limb usually do not affect space weather much at Earth. The major import of magnetic regions from the Sun's far hemisphere is that solar rotation carries them into the near hemisphere within two weeks, where their full impact on the Earth and near-Earth environment can be felt. As a single characteristic parameter, this two-week time scale by itself lends little sense of the realistic suddenness with which solar activity can elicit such an impact. Our star can gestate an active region from infancy to full maturity in its far hemisphere over a week or two to confront our planet with its direct impact from its east limb inside of an hour. Figure 1 illustrates an M-class flare of solar cycle 25 that emanated suddenly from just behind the far hemispheric east solar limb from an active region (NOAA AR12781) born in the far hemisphere in the previous solar rotation ([Jain 2020](#)), to inflict an unwelcome radio communication outage on southern Asia.

Synchronic observations of the full Sun are thus crucial for monitoring the emergence and evolution of its active regions, which are the primary drivers of space weather. Thus, both hemispheres provide vital information about solar activity that plays a decisive role in space weather forecasts. During the early rising phase of solar activity cycle 25, large active regions (NOAA AR 12785/12786) emerged on the farside during a relatively quiet period. This active region complex appeared in the farside seismic monitors and then was forecasted successfully about 10 days prior to its appearance on the frontside ([Jain & Lindsey 2020](#)). These did not produce any major geomagnetic storm but did increase the EUV flux ([Jain et al. 2021](#)). Figure 1 shows the complex crossing central solar meridian 18 days after its discovery in helioseismic maps of the Sun's far hemisphere.

The scope of space weather is about to expand radically over the next decade. The space weather we have known emanates from conspicuous magnetic regions in plain view but highly dependent upon location. In early February of 2022, 38 Starlink satellites were lost within a couple of days of their launch due to a geomagnetic storm originating from a CME driven by an M1.4-class flare emanating from NOAA AR12936. About two weeks later, the same active region produced another strong CME but on the farside, which was captured by several spacecrafts, e.g., the *Solar*

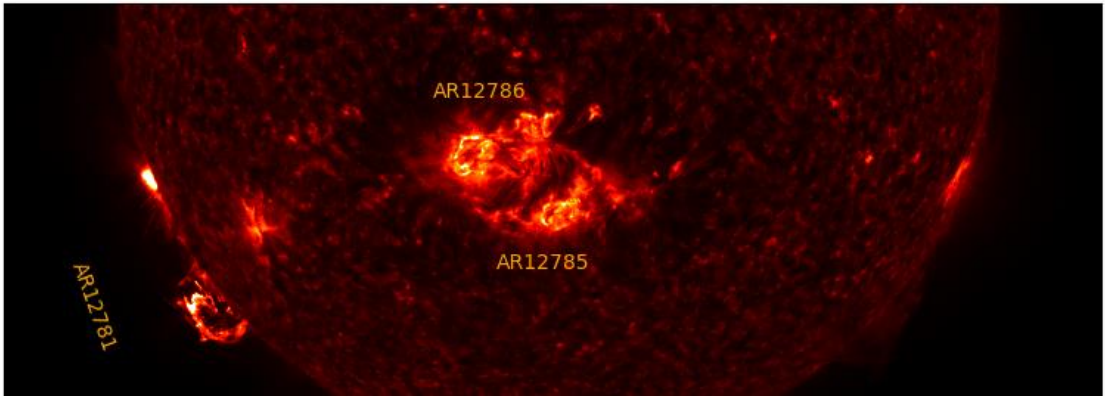

**Figure 1:** *SDO/AIA image of NOAA AR12781 in He II λ304- Å radiation releasing an M4.7-class flare from the Sun's east limb on 2020-11-29 after a full solar rotation since its birth in the east far hemisphere. Once the active region began its second transit of the near hemisphere, NOAA redesignated it AR12790.*

*Terrestrial RElations Observatory (STEREO)*, the *Solar Orbiter (SO)*, and the Large Angle and Spectrometric Coronagraph Experiment (LASCO) onboard the *Solar and Heliospheric Observatory* (*SoHO)*. Both CMEs hurled a significant amount of plasma and accompanying magnetic field into space. However, the farside eruption did not produce a geomagnetic storm at Earth, as the CME was directed away from us. It did, however, radically affect the interplanetary space environment near Mars. Monitoring of magnetic activity over the entire solar surface, 0° - 360° in Carrington longitude, then, is rapidly becoming an essential component of space weather forecasting in NASA's vision of human travel through interplanetary space.

Forecasting of space weather related parameters (e.g., solar wind, CMEs) depends on near-real-time data-driven magnetic field models ([Gonzi+2021](#); [Riley & Ben-Nun 2021](#)). Though the models (e.g., WSA-Enlil) have been successful to some extent for forecasting conditions in the space environment several days in advance, these are far from complete. The models are based on observations of the hemisphere facing the Earth, and the lack of information from the far hemisphere introduces severe gaps in the data stream with consequent errors in its forecasts. Over the past decade, some studies have been carried out to quantify the influence of data gaps on various space weather related parameters. For example, the case studies of backcasting solar wind and coronal holes by including farside information in the data assimilation provide encouraging results ([Arge+2013](#)). Synoptic magnetic flux maps from the Air Force Data Assimilative Photospheric Flux Transport (ADAPT; [Arge+2010](#)) model were used in the Wang-Sheeley-Arge (WSA; [Arge & Pizzo 2000](#)) model where the instantaneous global magnetic field was required for defining the inner boundary conditions to forecast solar wind. At present, ADAPT global maps are created with line-of-sight full-disk magnetograms only and neglect the active regions on the farside. [Arge+2013](#) have shown that the inclusion of farside active regions, particularly near limbs, can significantly refine the WSA model predictions by improving estimates of the global magnetic flux distribution. A snapshot of the results from this case study is presented in Figure 2. As illustrated, the solar wind forecast is significantly improved by including the farside active regions.

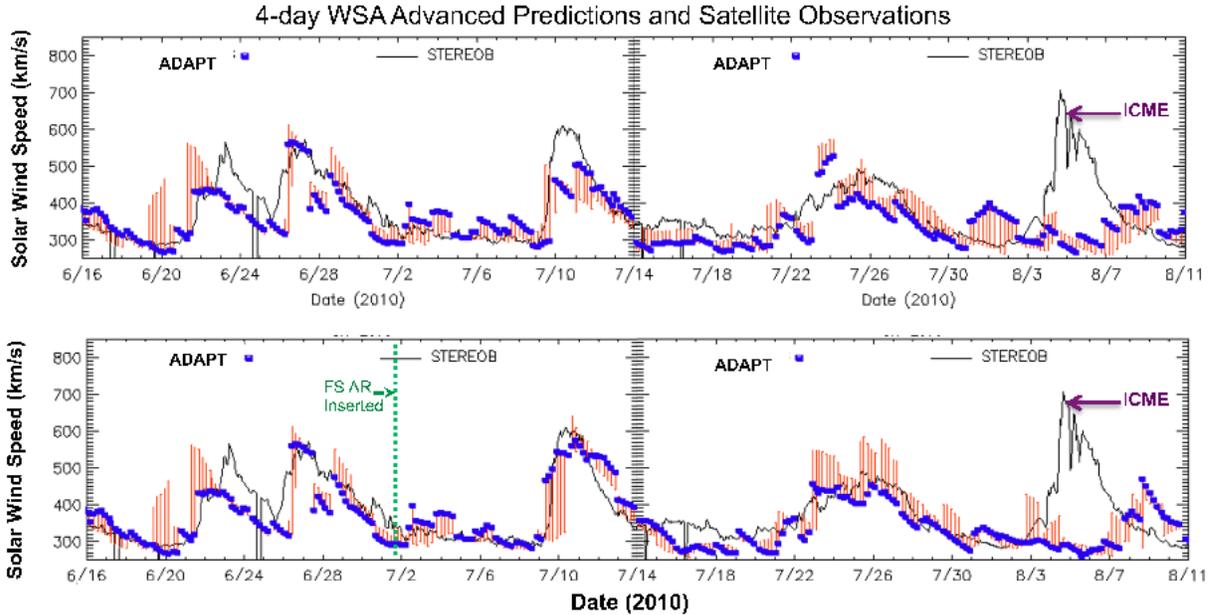

**Figure 2:** *Solar wind speed observations from STEREO B (black solid line) vs. 4-day WSA predictions (blue dots) using daily ADAPT maps without (top) and with farside active regions included (bottom). The red vertical bars indicate the range over which WSA solar wind speed predictions vary over a grid cell. Adapted from [Arge+2013](Arge+2013). The inclusion of farside information clearly improves the forecast.*

Another example is also based on the ADAPT-driven empirical model for forecasting the 10.7-cm radio flux. In this case, the current version of ADAPT provides estimates of the flux values on a time scale of 1 to 7 days in advance. However, the inclusion of solar farside magnetic activity estimates, i.e., the emerging and decaying fields, not only improves the forecast values but also extends forecasts beyond 7 days ([Henney+2012](Henney+2012)). Further, it has been demonstrated that the short-term forecast of UV irradiance can be improved by adding farside seismic maps ([Fontenla+2009](Fontenla+2009)).

The forecasting of flares, which can adversely impact the lower ionosphere, is not yet performing well near the east and west limbs. Due to the lack of information on active regions' photospheric magnetic morphology or evolutionary history that is hidden from view during farside transit, flare forecast methods routinely fail to reliably predict events that occur near or just beyond the solar limbs ([Park+2020](Park+2020)). Though flux transport models have been used to estimate the magnetic flux from the active regions after crossing the west limb, the situation is not so good near the east limb. The farside information becomes critical if an active region grows significantly or a new strong region emerges on the farside. By including information from the farside seismic maps in the flux transport models, one can quantitatively enhance the ability to describe farside solar magnetic activity. Finally, these enhanced capabilities can be used for predicting solar flares. As mentioned before, flares are related to CMEs and SEPs, hence it is not just the near Earth environment that can be impacted by the flares, but the entire solar system. The total flux in an active region can arguably provide respectable flare forecasts ([Leka & Barnes 2007](Leka+Barnes+2007); [Barnes+2016](Barnes+2016)). Incorporating estimates of total flux should provide advanced warning of near-limb but still Earth-impactful energetic events. Flare forecasts for the full Sun can be made with just rudimentary information from the farside seismology.

The advantages of farside imaging are not limited to operational space weather. These observations are critical for developing a comprehensive understanding of full life cycles of the magnetic regions from the early stages of their emergence to their growth to potential sources of flares, and to their eventual decay. With lifetimes up to several months, most of this evolution is unobservable from the Earth perspective alone. This can now be secured by continuous, stable, reliable synoptic monitoring of the Sun's far hemisphere over the coming decade.

## 3. **Current Status of Farside Observations and Mapping**

Direct images of the entire Sun's farside became available a few years after NASA launched *STEREO*, in October of 2006, with two nearly identical space-based observatories - one ahead of Earth in its orbit (*STEREO-A)* and the other trailing behind (*STEREO-B)*. *STEREO* provided full coverage of the Sun's far hemisphere in a few extreme-ultraviolet (EUV) wavelengths for several years, starting in 2011, until contact with one of its satellites was lost in October of 2014. Since then, only a limited portion of the farside has been observed by *STEREO*, depending upon the location of its still operational spacecraft (*STEREO-A*) in heliocentric orbit. Direct EUV coverage of the Sun's farside is diminishing day-by-day as *STEREO-A* is now progressing ever further back into the Earthside of the solar system, to leave the far hemisphere completely in the dark in about a year, before beginning its second passage towards the backside. However, due to the loss of *STEREO-B*, the coverage of the farside will be limited in the future years. *Solar Orbiter*, launched in February of 2020, will provide observations of the farside, including the first-ever images of the Sun from a high-latitude vantage. However, like that of *STEREO*, coverage of the farside by *Solar Orbiter* will be partial and intermittent. This mission is rather planned to observe one polar region at a time for relatively short periods; hence, it cannot realistically provide the continuous, stable, synoptic-quality monitoring needed for active-region research and space-weather forecasting in the coming decade.

Indirect methods involve mapping of the farside by the techniques of local helioseismology. Farside helioseismic maps are being routinely created using the frontside Doppler observations from the NSF-funded ground-based Global Oscillation Network Group (GONG) as well as the NASA-funded space-borne Helioseismic and Magnetic Imager (HMI) onboard the *Solar Dynamics Observatory (SDO)*. These maps are based on the principle that there is a phase shift (travel time delay) between waves entering and exiting a region of high magnetic field concentration. Farside maps were first produced by applying the technique of helioseismic holography to Michelson Doppler Imager (MDI)/*SoHO* observations ([Lindsey & Braun 2000](#); [Braun & Lindsey 2001](#)). This technique is now being used to compute maps for both [GONG](#) and [HMI](#) observations. The validity of the farside maps have been checked using *STEREO* EUV images for limited periods where about 90% of the large active regions' identification from seismic mapping coincided with those in the *STEREO* maps ([Liewer+ 2012](#), [2014](#)). Furthermore, over the past two decades, helioseismic mapping has progressed significantly ([Lindsey & Braun 2017](#)). The seismic signature in the farside maps as represented by phase shifts have also been calibrated against the magnetic field strength ([González Hernández+2007](#)); however, it needs to be revisited with concurrent seismic and magnetic field maps. The technique of time distance was also applied for farside mapping ([Duvall & Kosovichev 2001](#); [Zhao 2007](#)). It has been recently improved (Zhao+2019) and became operational on HMI observations. We show in Figure 3 examples of combined frontside magnetograms and farside seismic maps from the GONG pipeline. These maps display the magnetic activity in full 360° Carrington maps of the solar surface. The evolution of two active regions (marked by the white arrows) is accurately tracked in these composite maps. An inspection of the farside maps from all three pipelines (not shown here) displays a close agreement between them irrespective of the technique or data used. At present, the farside maps are routinely produced in near-real-time at a cadence of 12 hours, providing important information about the distribution and sizes of active regions in the far hemisphere. As discussed in the next section, farside mapping needs to be improved with additional capabilities for space weather forecasting.

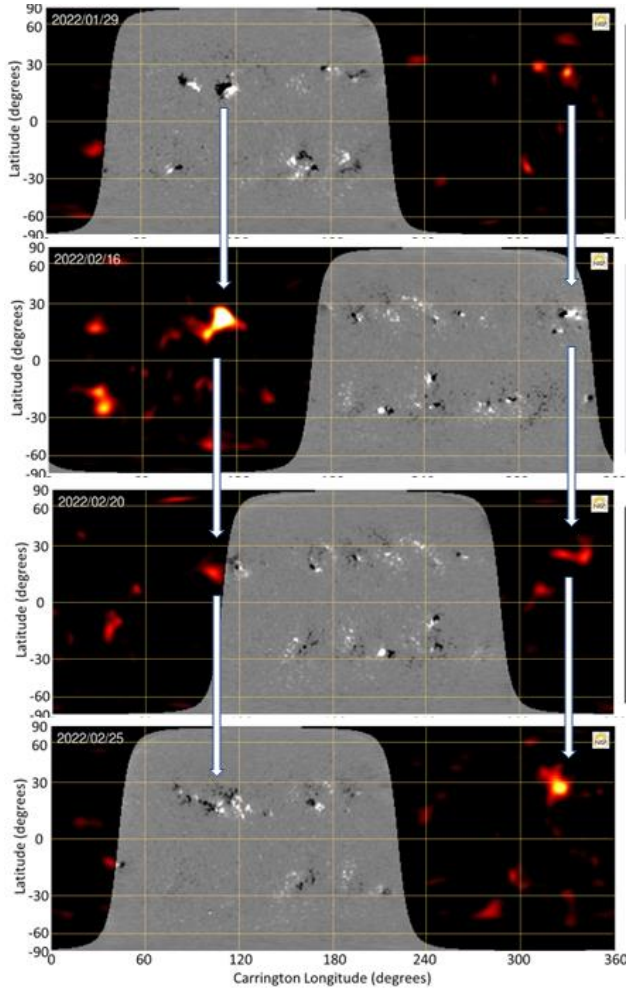

**Figure 3**: *Carrington maps of 360° Sun by combining farside helioseismic maps and frontside magnetograms for four days during 2022 Jan 29 and Feb 25. These are derived from the GONG observations with helioseismic holography – frontside and farside are represented by grayscale and black background, respectively. In each map, two long-lived active regions, NOAA 12936 (left) and NOAA 12941 (right), are marked by the white vertical lines.*

## 4. Looking Ahead: Needs and Opportunities for Space Weather Forecasting in the Coming Decade

The elementary requirement for precise space weather forecasting is to assimilate magnetic field information from both near- and far-hemispheres into our near-real-time data-driven magnetic-field models. This requires coordinated efforts by observers, data analysts and modelers. While numerous resources are underway to observe the front hemisphere at different wavelengths and cadences, active regions in the far hemisphere need to be acquired differently, with indirect methods (e.g., helioseismic mapping) until we have stable, direct observations of the Sun's farside as we presently have for the nearside.

The helioseismic mapping in operation today has evolved significantly since its inception. However, it may be further refined by improving the sensitivity of seismic signatures and improved models of their associated magnetic flux distributions. This will allow us to better recognize active regions on the farside and more accurately project conditions and timing of their appearance into the nearside, thus enabling better characterization of EUV irradiance and flaring potential.

Below we summarize some areas where additional work is needed for making farside maps more suitable for space weather forecasting,

- Develop higher spatial resolution in farside maps with reduced noise.
- Improve the sensitivity of helioseismic signatures to magnetic flux in the far hemisphere.
- Employ the Hale polarity law to helioseismic maps (see [Henney+2011](#)) to model the signed magnetic flux density of active regions from helioseismic signatures.
- Model magnetic inclinations in active regions in the far hemisphere.
- Develop EUV source proxies of helioseismic signatures.
- Robust identification of active regions with the application of machine learning tools. Several efforts are currently being conducted to achieve this goal (e.g., [Felipe & Asensio Ramos 2019](#), [Broock+2021](#)).

Furthermore, for reliable farside maps using helioseismic observations, a high duty cycle (approximately 85%) is required for good signal-to-noise. While this can be usually achieved from space, observations from ground-based observatories are subject to terrestrial weather and the diurnal cycle. However, a recent study based on the 18 years of GONG observations shows that such a high duty cycle can be achieved with a ground-based network of 6 sites ([Jain+2021](#)). The reported mean duty cycle in this study is 93.5%, which is similar to instruments placed on Low Earth Orbit (LEO) due to periodic eclipses. For comparison, HMI loses up to about 3.1 hr of observations, i.e., 13% of its 24-hr phase correlation in its longest terrestrial eclipses of the Sun, leaving a duty cycle of 87%. This does not heavily diminish the sensitivity of the resulting helioseismic signatures.

For improving space weather forecasts, efforts are needed to upgrade the existing models by incorporating both nearside and farside active regions in near-real time. This includes refinement in WSA-Enlil/ADAPT modeling being used by NOAA/SWPC to provide 1–4 day advance warning of solar wind structures and Earth-directed CMEs as well as by Space Environment Technologies' solar proxy forecasts for estimating irradiance variations on timescales of a few days that are relevant to re-entry of deorbiting debris. This could include major development of an empirical scheme that would employ the Hale polarity law to model the magnetic polarities of different components of helioseismic signatures, which by themselves are insensitive to the sign thereof ([Henney+2011](#)).

The full-sun boundary data is required for global Potential-Field Source Surface (PFSS) maps that form initial conditions for many heliospheric models. In these models, farside magnetic maps are constructed in various ways invoking several assumptions; the only way at present to mitigate these assumptions is through high-sensitivity high-accuracy farside seismic mapping.

In summary, the advantages of incorporating farside seismic maps in space weather forecasting include the following:

- Prior information about newly emerged active regions in the far hemisphere and their impending arrival into the frontside.
- Improved constraints on data-driven magnetic field models.

- Better understanding of solar phenomena that produce severe space weather events.
- The ability to assess and correct systematic errors in forecasting algorithms.
- Extended prediction lead times.
- Improved warnings of extreme solar weather events, e.g., flares near the limbs.
- Better EUV irradiance forecasts.
- To foster scientific collaboration between observers, data analysts, modelers, and space weather forecasters.

It is important to note that both the existing data sources for farside seismic mapping now rely on aging instruments. GONG has been operational for 27 years, and SDO is now in its 13th year. If these instruments fail, the Sun's far hemisphere will fall largely back into the dark. No future space mission has been planned/finalized so far by any space agency to provide full, stable, continuous coverage of the Sun's far hemisphere. There have been some on-going efforts towards positioning spacecrafts at the L4/L5 Lagrange points (e.g., The Multiview Observatory for Solar Terrestrial Science- MOST; Gopalswamy+2021 and several white papers submitted to this survey). These spacecrafts extend the coverage by 60° behind the west or east limb, respectively.

Therefore, in the absence of any planned space mission suitable for farside observations or mapping, the existing instruments at GONG facilities need to be upgraded with improved capabilities. The National Solar Observatory and the High Altitude Observatory are working together on a proposed upgraded network, currently known as next generation GONG (ngGONG, see White Paper by Pevtsov+ 2022, *ngGONG - Future Ground-based Facilities for Research in Heliophysics and Space Weather Operational Forecast*). ngGONG observations will not only provide continuity to the data required for farside mapping, but its multi-height observations will also offer additional improvements to farside mapping (see White Paper by Tripathy+2022, *Improving the Understanding of Subsurface Structure and Dynamics of Solar Active Regions*). These observations will further be used in predicting space weather on intermediate time-scales (see White Paper by Dikpati+2022, *Space Weather Modeling and Prediction for Intermediate Time-Scales*).

## 5. Observational Requirements for Helioseismic Farside Mapping

For continuous monitoring of the Sun's far hemisphere, uninterrupted full-disk Doppler observations at a cadence of 45 s (similar to HMI) or 60 s (similar to GONG and MDI) are required. In addition, the spatial resolution should not be less than 2″ per pixel. All these requirements are within the specifications of the proposed ngGONG.

## 6. Recommendation

Helioseismic farside imaging is a valuable technique for producing a complete picture of solar activity including the side of the Sun that cannot be directly observed from the Earth. Continuous, reliable synoptic monitoring of the Sun's far hemisphere offers a much-needed information for our world-wide needs in fundamental solar research, including operational space weather forecasting in the 21st century. This needs to be continued and further developed as an integral component of our science and technology infrastructure over the coming decade.